\documentclass{aa}  
\usepackage{graphicx}
\usepackage{txfonts}
\begin{document}
\newcommand{\rmn}[1] {{\rm #1}}
\newcommand{\Zsolar}{\mbox{${\; {\rm Z_{\sun}}}$}}
\newcommand{\ha}{\hbox{H$\alpha$}}
\newcommand{\oii}{\hbox{[O\,{\sc ii}]}}
\newcommand{\neiii}{\hbox{[Ne\,{\sc iii}]}}
\newcommand{\siii}{\hbox{[S\,{\sc iii}]}}
\newcommand{\ariii}{\hbox{[Ar\,{\sc iii}]}}
\newcommand{\nii}{\hbox{[N\,{\sc ii}]}}
\newcommand{\sii}{\hbox{[S\,{\sc ii}]}}
\newcommand{\oiii}{\hbox{[O\,{\sc iii}]}}
\newcommand{\hb}{\hbox{H$\beta$}}
\newcommand{\hg}{\hbox{H$\gamma$}}
\newcommand{\hd}{\hbox{H$\delta$}}
\newcommand{\hi}{\hbox{H\,{\sc i}}}
\newcommand{\hii}{\hbox{H\,{\sc ii}}}
\newcommand{\heii}{\hbox{He\,{\sc ii}}}
\newcommand{\etal}{\hbox{et\thinspace al.\ }}
\newcommand{\oiiha}{\hbox{[O\,{\sc ii}]/H$\alpha$}}
\newcommand{\fha}{\hbox{$F_{{\rm H}\alpha}$}}
\newcommand{\fhb}{\hbox{$F_{{\rm H}\beta}$}}
\newcommand{\foii}{\hbox{$F_{\rm [O\,{\sc II}]}$}}
\newcommand{\micron}{\hbox{$\mu$m}}
\newcommand{\zoh}{\hbox{$12\,+\,{\rm log(O/H)}$}}
\newcommand{\teoii} {\hbox{$T_e{\rm (O\,{\sc II})}$}}
\newcommand{\tenii} {\hbox{$T_e{\rm (N\,{\sc II})}$}}
\newcommand{\teoiii}{\hbox{$T_e{\rm (O\,{\sc III})}$}}

\authorrunning {Shi \etal}
   \title{ \neiii/\oii \  as an oxygen abundance indicator in the $\hii$ regions and $\hii$ galaxies}


   \author{
          F. Shi\inst{},
          G. Zhao\inst{},
          Y. C. Liang\inst{}
          }

   \offprints{F. Shi}

   \institute{
              National Astronomical Observatories, Chinese Academy of Sciences, 20A Datun Road, Chaoyang District, Beijing 100012, PR China\\
              \email{fshi@bao.ac.cn}
             }

   \date{Received 29/01/2007; accepted 24/08/2007}

 
  \abstract
   {}
   {
    To calibrate the relationship between  $Ne3O2$ 
  ($Ne3O2$=log(\neiii$\lambda3869$/\oii$\lambda3727$)) and
  oxygen abundances,    we present a sample of $\sim$3000 $\hii$
  galaxies from the Sloan Digital Sky Survey (SDSS)  
    data release four. They are  associated with a sample from the literature 
  intended to  enlarge the oxygen abundance region.
   }
   {
¡¡  We calculated the electron temperatures ($T_e$) of 210 galaxies in the SDSS sample 
   with the direct method, 
   and  $T_e$ of the other 2960 galaxies in SDSS sample  calculated
   with an empirical method. Then, we use a linear least-square fitting 
   to calibrate the  $Ne3O2$ oxygen abundance indicator.  
     }
   { It is found that the $Ne3O2$ estimator follows a linear relation 
   with \zoh \
   that holds for the whole abundance range covered by the sample, from 
   approximately 7.0  to 9.0. The best linear relationship between the 
    $Ne3O2$  and the oxygen abundance is calibrated. The dispersion between oxygen abundance and $Ne3O2$   index in the metal rich
    galaxies may come partly from   the moderate depletion of oxygen onto grains.
    The $Ne3O2$ method has the virtue of being single-valued and 
    not affected by internal reddening. As a result, 
   the $Ne3O2$ method can be a good metallicity indicator in the $\hii$ regions and $\hii$ 
     galaxies, especially  in  high-redshift galaxies.  
   }
  {}

\maketitle
%

\section{Introduction}

Those hydrogen II galaxies with strong emission lines are important probes for
the formation and evolution of galaxies. Their spectra contain the much
important information needed to determine the star formation
rate, initial mass function, element abundance, etc. (Stasi{\'n}ska \&
Leitherer 1996; Kennicutt 1998; Contini \etal 2002).
The heavy element abundance is a key parameter for the
formation and evolution of a galaxy. Oxygen is the 
important element that is easily and reliably determined since
all the most important ionization stages can  be observed.
The oxygen abundance from the measurement of electron temperature
from  \oiii$\lambda\lambda$4959,5007/\oiii$\lambda$4363 is
one of the most reliable methods.
But \oiii$\lambda4363$ is usually weak in the low metallicity galaxies,  
and there are often large errors when measuring this line. In high 
metallicity galaxies, \oiii$\lambda4363$ are hardly even observable.
 
Instead of the $T_e$ method, strong line methods,  such as the $R_{23}$ 
 \footnote{$R_{23}$=(\oii$\lambda$3727+\oiii$\lambda\lambda$4959,5007)/\hb}, 
$P$ \footnote{$P$=\oiii$\lambda\lambda$4959,5007/(\oii$\lambda$3727+\oiii$\lambda\lambda$4959,5007)}, 
$N2$ \footnote{$N2$=log(\nii$\lambda$6583/\ha)}, or 
$O3N2$ \footnote{$O3N2$=log((\oiii$\lambda$5007/\hb)/(\nii$\lambda$6583/\ha))} methods, are  used widely 
 (Pagel \etal 1979; Kobulnicky, Kennicutt, \& Pizagno 1999; Pilyugin \etal 2001; Charlot
\& Longhetti 2001; Denicol{\'o} \etal 2002; Pettini \& Pagel 2004;
Tremonti \etal 2004; Liang \etal 2006). The $R_{23}$ and  $P$ methods suffer the 
double-valued problem, requiring some assumption or rough  a priori 
knowledge of a galaxy's metallicity in order to locate it on the 
appropriate branch of the relation. The $N2$- and  $O3N2$ methods
are monotonic, but the reasons for this are not purely  physical. It is  
partly due to  the N/O ratio increases on average with the increase in  
metallicity (Stasi{\'n}ska 2006; Shi \etal 2006). Besides, calibrations 
of the $O3N2$ and $N2$ indices might be improper for interpreting the integrated 
spectra of galaxies because $\nii\lambda6583$  and \ha\ may arise not only in bona 
fide $\hii$ regions, but also in the diffuse ionized medium.
Stasi{\'n}ska (2006) has recently proposed $Ar3O3$ \footnote{$Ar3O3$=\ariii$\lambda7135$/\oiii$\lambda5007$}  and $S3O3$ \footnote{$S3O3$=\siii$\lambda9069$/\oiii$\lambda5007$} as new abundance 
indicators, which have the advantage of being unaffected by the 
chemical evolution effects. The advantages are superior to previous $N2$  and $O3N2$ 
methods. 
 
In short, one ideal metallicity indicator has to be monotonic and be independent 
of the internal reddenning and chemical evolution. 
Nagao \etal (2006) find that the $Ne3O2$ index, i.e. 
log($\neiii\lambda3869/\oii\lambda3727$),  fulfills these requirements.  They  
derive a relation of $Ne3O2$ $vs.$ \zoh\ by using the Bayesian abundances provided
by the MPA/JHU group \footnote{See http://www.mpa-garching.mpg.de/SDSS/.} for 
the metal rich galaxies in the SDSS (Tremonti \etal 2004) and the $T_e$ abundances based on 
 an electron temperature-sensitive line, $\oiii\lambda4363$, for metal poor galaxies. The
 Bayesian metallicity could be  problematic in some aspects, 
such as improper subtraction of  the underlying starlight or  unsuitable treatment of nitrogen enrichment in the HII galaxy model. As a result, there is a significant systematic difference between Bayesian metallicity and $T_e$ metallicity (Yin \etal 2007),  
 so it is necessary to recalibrate the $Ne3O2$ index based on
 abundance derived from $T_e$ method, which is believed to be the most
 reliable method to determine oxygen abundance. Also, the dispersion
 between $\neiii\lambda3869/\oii\lambda3727$ and oxygen abundance needs to be explained.

In this paper, we recalibrate the $Ne3O2$ metallicity index based on a large 
sample of $\hii$ regions and $\hii$ galaxies where oxygen abundance is determined 
by the $T_e$ method.  For low metallicity galaxies (Sample I, see Sect.2), 
we used a five-level statistical equilibrium
model in the IRAF NEBULAR package (de Robertis \etal 1987;
Shaw \& Dufour 1995), which makes use of the latest collision
strengths and radiative transition probabilities to determine the
$T_e$ and $n_e$. For high  metallicity galaxies (Sample II, see Sect.2), an
empirical relation of $T_e$ and strong spectral lines from Pilyugin (2001) was adopted
for the electron temperature determination (their Eq.11). 
We also  study the physical origin of 
the dispersion between $\neiii\lambda3869/\oii\lambda3727$ and oxygen abundance.

The Sloan Digital Sky Survey (SDSS) is  the  most ambitious imaging and spectroscopic 
survey  to date, and it will eventually cover a quarter of the sky (York et al. 2000).
The large area coverage  and moderately deep survey limit of the
SDSS makes it suitable for studying the evolution of galaxies. Because of its 
homogeneity, the SDSS provides   a large sample of $\hii$ galaxies where oxygen 
abundance can be calculated with the classic $T_e$ method. The sample can be used
to calibrate some metallicity indicators.

This paper is organized as follows. Based on an SDSS DR4 starbursts spectral sample
and a sample of $\hii$ regions or $\hii$ galaxies from the literature,
 we  present a  sample to use for our metallicity index 
calibration (Sect.\ref{sam}). In Sect. \ref{calibrator}, we calibrate the $Ne3O2$ 
metallicity index. In Sect.\ref{disc}, we study the origin of dispersion between
oxygen abundance and $Ne3O2$ index and check the accuracy of our calibration.
In Sect. \ref{conclusion}, we summarize the calibration result 
and discuss the merits of then $Ne3O2$ indicator with respect to the
 other strong line indicators and then conclude this paper. 

\section{Data sample}\label{sam}

The preferred method for determining  oxygen abundances in
galaxies is  obtained from the ratio of auroral to nebular line intensities,
such as \oiii$\lambda\lambda$4959,5007/\oiii$\lambda$4363 (the
so-called $T_e$ method). In this
paper, the adopted metallicities of $\hii$ regions and $\hii$ galaxies for 
calibration are determined from the $T_e$ method. 

We used $\hii$ galaxies  from the Fourth Data Release (DR4) of
 the SDSS. After subtracting the underlying
 starlight using the method of Li et al. (2005) and  Lu et al. (2006),
we fit  the emission line using the method of Dong et al. (2005).
We made the internal reddening correction for the flux of all the emission 
line, using the two strongest Balmer lines, \ha/\hb\ and 
the effective absorption curve
$\tau_\lambda=\tau_V(\lambda/5500{\rm\AA})^{-0.7}$, which was
introduced by Charlot \& Fall (2000). Then, we made use of the spectral
diagnostic diagrams from Kauffmann et al. (2003) to classify galaxies
as starburst galaxies, active galactic nuclei (AGN),
or unclassified.  To reduce systematic and random errors from aperture 
effects, our galaxy samples are limited by the requirement that  
redshift $z>0.04$ (Kewley \etal 2005).

Within the primary starburst sample, two subsamples  were selected 
from the SDSS-DR4 with the fluxes of spectral
lines for all $\neiii\lambda3869$, \oii$\lambda3727$, \hb,
\oiii$\lambda4959$, \oiii$\lambda5007$, \ha\, and \nii$\lambda6583$  
higher than 5 times the flux uncertainty. The difference of these
two subsamples is that the first subsample ({\bf Sample I}) was selected by the additional
criterion from the \oiii$\lambda4363$ line, and the flux uncertainties
for \oiii$\lambda4363$ was higher than $5\sigma$. In all, 210 galaxies were
included in this subsample. The \oiii$\lambda$4363 is strongly
dependent on the metallicity of galaxies and it becomes undetectable
in high metallicity galaxies. Therefore, galaxies in Sample I are
those with low metallicity. The electron temperature in the 
high-temperature zone (\teoiii) of Sample I galaxies was calculated from \oiii$\lambda\lambda$4959,5007/\oiii$\lambda$4363, 
using a five-level statistical equilibrium
model in the IRAF NEBULAR package (de Robertis \etal 1987;
Shaw \& Dufour 1995). In the second subsample ({\bf Sample II}), 
galaxies have weak or no \oiii$\lambda$4363 line, and 2960 galaxies were 
selected in this subsample containing generally metal rich galaxies. 
The \teoiii\ of the galaxies in Sample II was calculated with the empirical method 
following Pilyugin (2001) from the $R_{23}$ and $P$ parameter (their Eq.11).

To estimate the temperature in the low-temperature zone \teoii, the 
relation between \teoii\ and \teoiii\ from Garnett (1992) are  
utilized : 
\begin{equation} 
t_e{\rm (O\,{\sc II})} = 0.7 \times t_e{\rm (O\,{\sc III})} + 0.3, 
\end{equation} 
where $t_e$=$T_e/10^4$ K. After calculation of $T_e$, we used the expressions from Izotov \etal (2006) 
 (their Eqs.3 and 5) to calculate the oxygen abundances. 
The  average uncertainty from 
flux measurement in the computed \zoh\ values is typically 
0.10 dex in both samples.

To widen the oxygen  abundance range, we built a large 
database of published reddening-corrected line fluxes for
$\hii$ regions or $\hii$ galaxies, besides the $\hii$ galaxies from SDSS DR4. 
 The  data for $\hii$ regions in spiral
galaxies in the literature with good $\neiii\lambda3869$,
$\oii\lambda3727$ and reliable $T_e$ metallicity are taken from
Garnett \& Kennicutt (1994), Garnett \etal (1997),
Bresolin \etal (2004), Bresolin \etal (2005) and van Zee \etal (1998).
The  data for $\hii$ galaxies in the literature with very good
signal-to-noise ratios are taken from
Guseva et al. (2003a,2003b,2003c), Izotov et al. (1994, 1997, 1998, 2004a, 2004b, 2006),
Lee et al. (2003a, 2003b), Lee et al. (2004), Melbourne et al. (2004),
Papaderos et al. (2006), and Thuan et al. (1995). In total,  
84 $\hii$ regions of spiral galaxies and 446 $\hii$ galaxies in 
the literature are adopted in our sample.
The average uncertainty from flux measurement in the \zoh\ values 
is typically within 0.10 dex. 
\section{$Ne3O2$ as a metallicity calibrator}\label{calibrator}

\begin{figure}
\centering
\includegraphics[angle=0,width=0.45\textwidth]{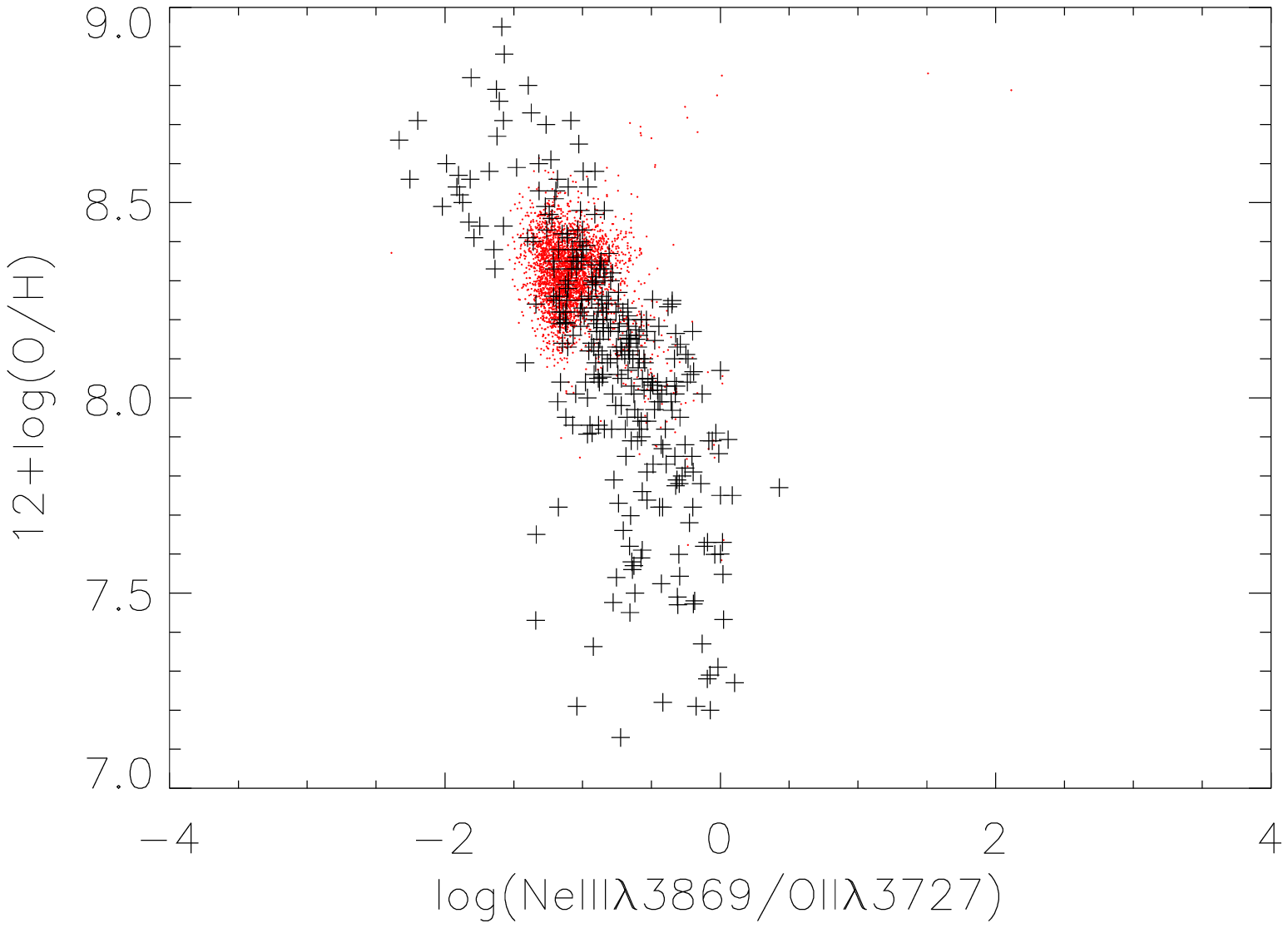}
\includegraphics[angle=0,width=0.45\textwidth]{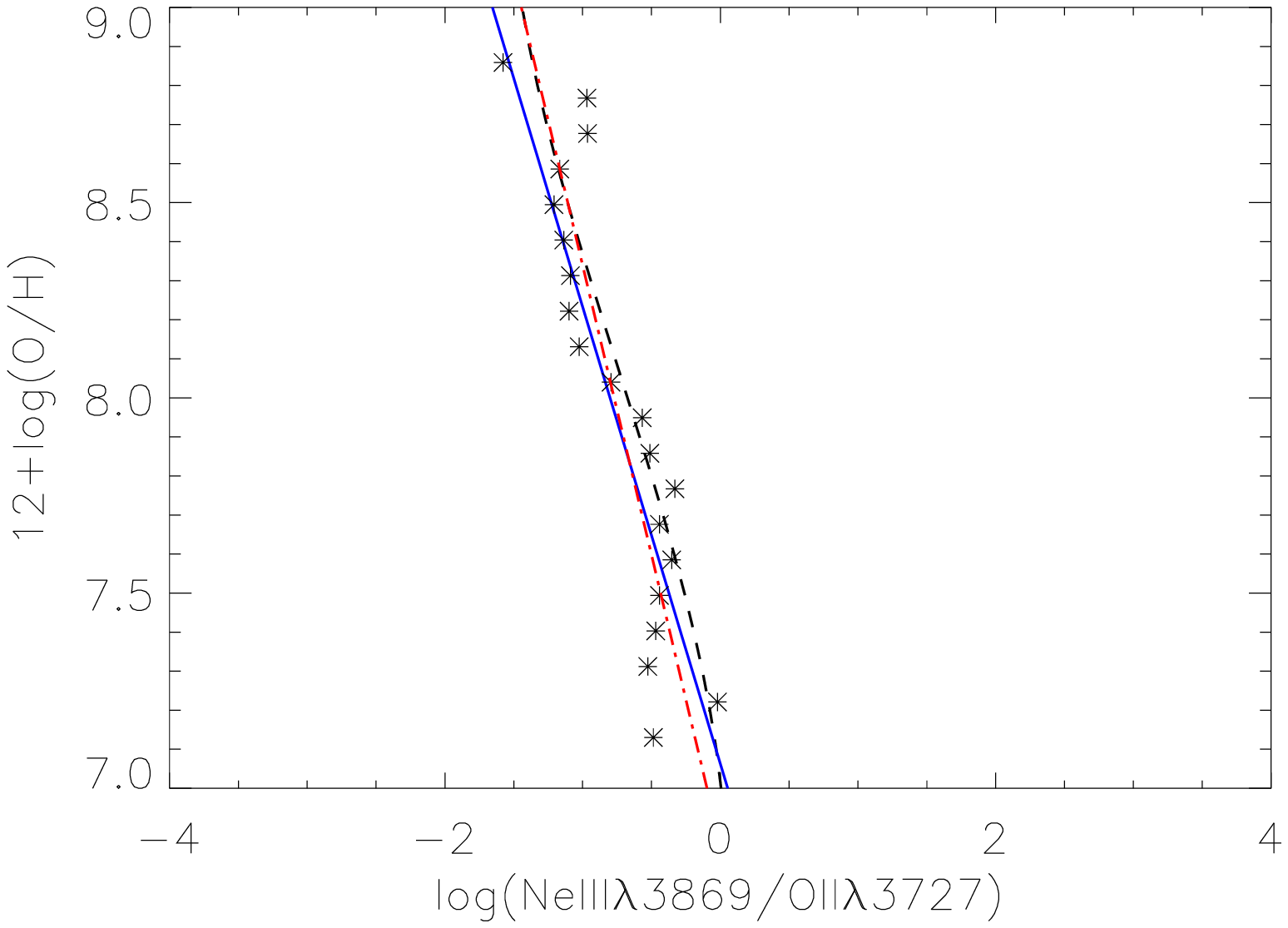} 
\caption{ 
Upper panel:  $\zoh$  vs. log(Ne3O2) for our data sample. 
Dots are galaxies from SDSS DR4. Crosses are from the literature. 
Lower panel: The solid line represents Eq. (2). Stars denote the mean 
 for each bin of the oxygen abundance. 
The dot-dashed lines denote the best-fit  function for each metallicity bin. 
The dashed curves are the calibration of  Nagao \etal (2006). 
 }
\label{metal-calibrator}
\end{figure}

The upper panel of Fig. \ref{metal-calibrator} shows the oxygen abundance as a function of
 $Ne3O2$ for our samples.  Apart from a few outliers in the SDSS DR4 sample,  all 
the objects merge into a relatively tight, linear, and steep sequence, 
as expected. As expected from stellar nucleosynthesis (Izotov \& Thuan 1999) and confirmed by
 observation of low metallicity  HII galaxies in the SDSS DR3 (Izotov \etal 2006),
 Ne and O are produced by the same stars, so the 
relationship between oxygen abundance and the $Ne3O2$
index unlike $O3N2$ or $N2$ is not affected by the chemical evolution effects. The observed trend
of $Ne3O2$ is due to the fact that the ionization parameter has a strong 
metallicity dependence (Nagao \etal 2006). The two emission lines, 
 $\neiii\lambda3869$ and $\oii\lambda3727$, have different degrees of ionization. 
 Their ratio is strongly dependent on 
the ionization parameter and correlates with metallicity. 

\begin{figure}
\centering
\includegraphics[angle=0,width=0.45\textwidth]{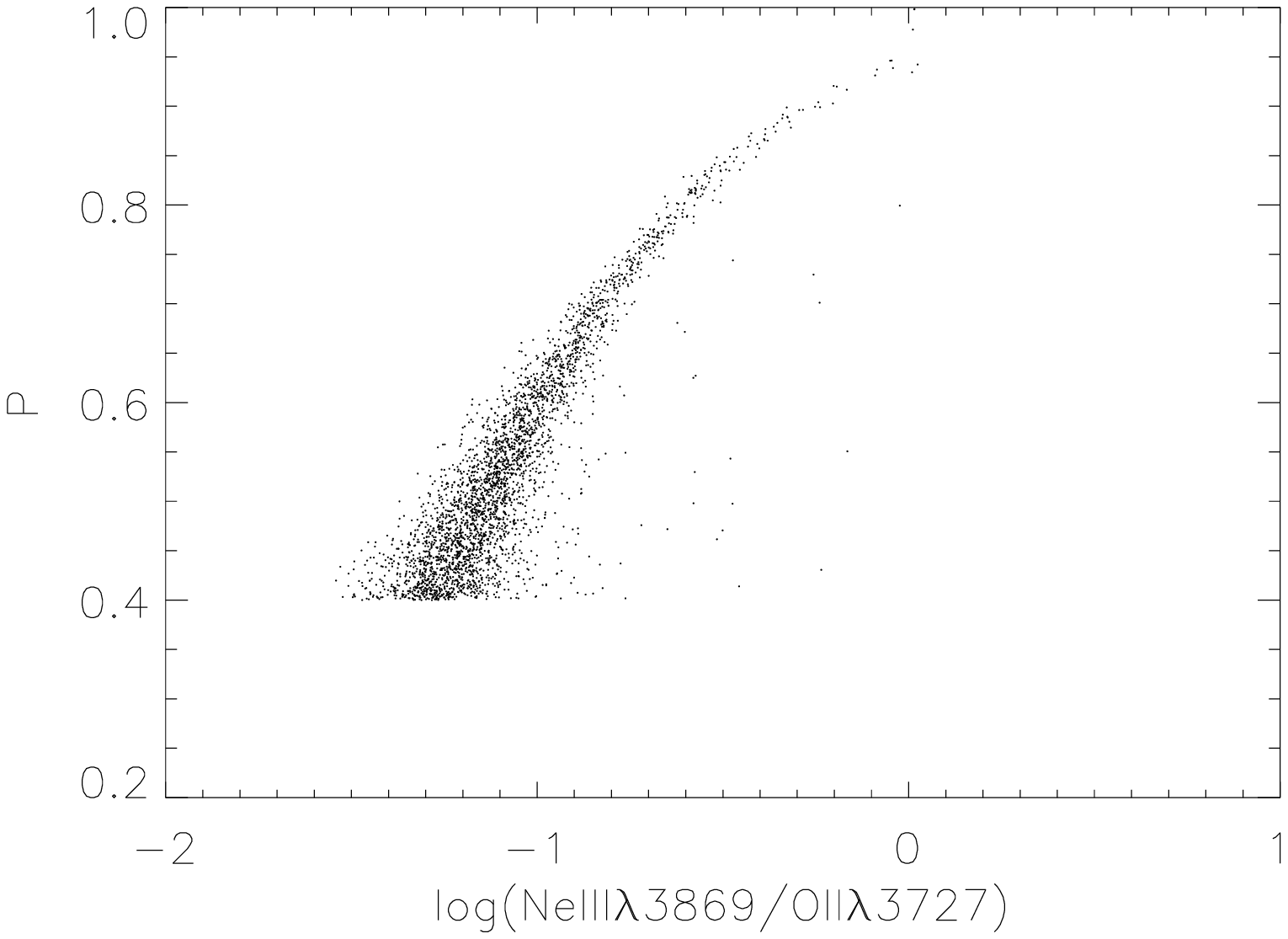}
\caption{ The relationship between $Ne3O2$ index and the P parameter for $\hii$ galaxies in SDSS DR4.
 P=(\oiii$\lambda4959$+ \oiii$\lambda5007$)/(\oii$\lambda3727$+\oiii$\lambda4959$+ \oiii$\lambda5007$)}
\label{p-ratio}
\end{figure}

\begin{figure}
\centering
\includegraphics[angle=0,width=0.45\textwidth]{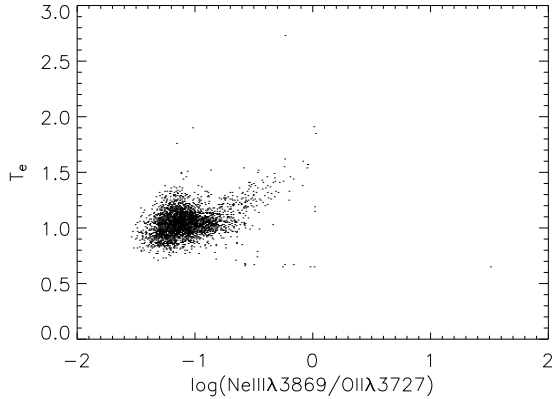}
\caption{ The relationship between $Ne3O2$ index and the electron temperature for $\hii$ galaxies in SDSS DR4. The electron temperature is in units of $10^4$K. }
\label{te-ratio}
\end{figure}

To confirm this view, we plot the relationship in Fig. \ref{p-ratio} between 
$Ne3O2$ index and the $P$ parameter for $\hii$ galaxies in SDSS DR4.  The $P$ parameter 
is defined as  (\oiii$\lambda\lambda4959,5007$)/(\oii$\lambda3727$+ 
\oiii$\lambda\lambda4959,5007$) by Pilyugin (2001), and it is a 
good  representive of an ionization parameter.  Figure \ref{p-ratio} shows that 
the $Ne3O2$ index is  positive related to the $P$ parameter, which supports the idea 
that the $Ne3O2$ index is negatively correlated with the metallicity,  since more metal rich
$\hii$ regions are excited  by a softer radiation  field and have a lower ionization
 parameter. We have  checked the relationship further between electron temperature 
$T_e$ and  $Ne3O2$ index in Fig. \ref{te-ratio} and find the observed trend of
$Ne3O2$  is  partly due to the increase in $T_e$ as metallicity 
decreases, which leads to an increase in the $Ne3O2$ ratio. The comparison
 between  Figs. \ref{p-ratio} and \ref{te-ratio} shows that, 
for the observed relation between \zoh\ and the $Ne3O2$ index, the 
contribution from the decrease in the  ionization parameter $P$  with 
the increase in metallicity is stronger than the decrease in $T_e$ 
with an increase in metallicity, since the  $Ne3O2$ index is more 
sensitive to the $P$ parameter than $T_e$.

From the upper panel of Fig. \ref{metal-calibrator}, we can define a new metallicity 
calibration. The observed distribution of all the points  in this figure 
is linear least-square-fitted by the following expression given as the
solid line in the lower panel of Fig. \ref{metal-calibrator}:

\begin{eqnarray}
\zoh =  -1.171(\pm0.088) \times Ne3O2  \nonumber \\ 
 + 7.063(\pm0.727)) 
\end{eqnarray}
with a standard error of  0.302 dex. In this panel, we also give the mean of the data with 
bin=0.091 dex in $\zoh$ (the stars) and the result of fitting the mean.  We plot the calibration of Nagao \etal (2006)  to compare with our result. In Sect.4.2, we discuss why we prefer the fit for all the points and do not use the fit 
of mean.

\section{Discussion}\label{disc}

\subsection{The origin of the dispersion }
\begin{figure}
\centering
\includegraphics[angle=0,width=0.45\textwidth]{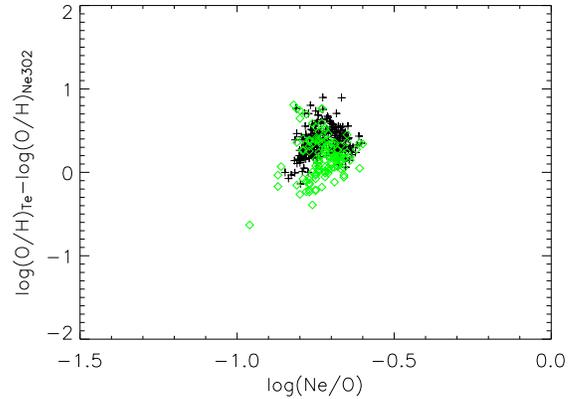}
\includegraphics[angle=0,width=0.45\textwidth]{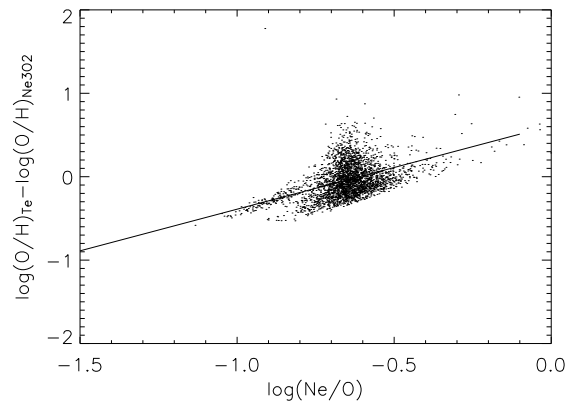}
\caption{ Upper panel: Correlation between  $\log(O/H)_{T_e}$-$\log(O/H)_{Ne3O2}$ 
and the $\log(Ne/O)$  abundance ratios of Sample I (crosses) 
and the sample of Izotov et al. (2006) (diamonds). 
 $\log(O/H)_{T_e}$-$\log(O/H)_{Ne3O2}$ is the difference between $T_e$ 
metallicity determined from the expressions in Izotov \etal (2006) 
and $Ne3O2$ metallicity determined from Eq.(2). 
Lower panel: Correlation between  $\log(O/H)_{T_e}$-$\log(O/H)_{Ne3O2}$ and the $\log(Ne/O)$
 abundance ratios of Sample II (dots). }
\label{dispersion}
\end{figure}

Although the $T_e$ metallicities are closely 
related to $Ne3O2$ index, the dispersion of the relation is large, especially 
in the low metallicity region. Some of the scatter may come from
the moderate depletion of oxygen onto grains in the metal rich galaxies.
To show this view clearly, we calculated the log(Ne/O) abundance ratios for 
{\bf  Samples I} and {\bf II} using the 
method of Izotov \etal (2006) (their equations (7) and (19)). Then we plotted  the 
difference between  $\log(O/H)_{T_e}$-$\log(O/H)_{Ne3O2}$ and
 the $\log(Ne/O)$ abundance ratios in the upper panel of 
Fig. \ref{dispersion} for metal poor galaxies and the lower panel for  
 metal rich galaxies. To compare {\bf  Sample I} and 
 metal-poor emission-line galaxies in Izotov et al. (2006), we  show the 
 sample of Izotov \etal (2006) in the upper panel of Fig. \ref{dispersion}. 

As expected, {\bf  Sample I} is consistent with the sample of 
Izotov \etal (2006). There is a positive correlation between 
 ($\log(O/H)_{T_e}$)-($\log(O/H)_{Ne3O2}$) and the log(Ne/O) 
for metal-poor galaxies because more oxygen depletion would be 
present in the $\hii$ regions with higher metallicity (Izotov \etal 2006 ).  

The lower panel of Fig. \ref{dispersion} shows more clearly that 
there is a clear correlation between $\log(O/H)_{T_e}$-$\log(O/H)_{Ne3O2}$ 
and the log(Ne/O) for the metal rich galaxies because more 
oxygen is locked in the dust grains in the more metal-rich 
$\hii$ regions (Izotov \etal 2006 ). This correlation for 
 {\bf  Sample II} can be given as a linear least-square fit:
\begin{eqnarray}
\log(O/H)_{Te}-\log(O/H)_{Ne3O2} =  0.610(\pm0.024)+ \nonumber \\
 0.999(\pm0.037) \times \log(Ne/O), 
\end{eqnarray}%
with a standard error of 0.104 dex. The corrected calibration of $Ne3O2$ methods by
considering this correlation  will provide more accurate oxygen abundance, 
and the standard error of the corrected calibration in Fig. \ref{metal-calibrator} 
will decrease to 0.198 dex, which is much less than the previous 
ones (0.302 dex). This correlation can be used to correct the derived 
oxygen abundances from $Ne3O2$ methods. However, to apply the Ne/O correction to 
12+log(O/H) abundances, 
we have to measure electron temperature $T_e$, \hb, 
\oiii$\lambda4959$, \oiii$\lambda5007$, besides  $\neiii\lambda3869$ and 
\oii$\lambda3727$, so it will not always be practical. 
 
\subsection{Comparison with others}

We plot the calibration of Nagao \etal (2006) to compare with Eq. (2) 
in the lower panel of Fig. \ref{metal-calibrator}. 
Some samples of both ours and Nagao's are selected randomly from 
the literature. As a result, the calibration might be affected by selection 
effects. To mitigate this effect, Nagao's calibration fits the mean in 
given metallicity bins.  
 However, most of the SDSS data in our 
sample are concentrated in the metallicity range of  8.0$<\zoh<$8.5. 
As a result, if we fit the mean, the calibration will lose a lot of  information
 in 8.0$<\zoh<$8.5, because thousands of SDSS data are  represented by several
 different mean values. As a result, the calibration fitting the mean will 
induce larger errors. In considering this, we used the calibration fitting 
 all data points in the lower panel of Fig. \ref{metal-calibrator}
 to compare with Nagao's calibration. 

The lower panel of Fig. \ref{metal-calibrator}  shows that the oxygen
 abundance from Nagao's calibration is  systematically higher  than our 
calibration at the given $Ne3O2$, except that at the low metallicity end, our 
calibration is consistent with Nagao's because both ours and Nagao's 
abundances at the low metallicity end are derived by the $T_e$ method. 
The difference between Nagao's calibration and ours in the high metallicity
 region could be caused by the oxygen abundances
 of most galaxies in the Nagao's sample being calculated by the Bayesian methods,  
while the oxygen abundances of our sample are all derived with the $T_e$ 
methods. The origin of the difference between the Bayesian 
 and $T_e$ metallicities have already been discussed by Yin et al. (2007).
 They find that, for almost half of the sample galaxies
 (227 among 531 galaxies with $T_e$ measurements), 
Bayesian  metallicities are overestimated by a factor of about 0.34 dex on average, 
 which is consistent with our result. They propose that the overestimates of
Bayesian metallicities may be related to the onset of secondary N enrichment 
in models.  Another reason for the  lower $T_e$ metallicities 
than Bayesian metallicities is that the \oiii$\lambda$4363 emission line is 
biased by the very hot $\hii$ regions in each galaxy; thus, the global 
average temperature might be overestimated to 1000-3000K, which results
 in systematic underestimation of the oxygen abundance of 
0.05-0.2 dex, as Nagao \etal (2006) propose.

\begin{figure}
\centering
\includegraphics[angle=0,width=0.45\textwidth]{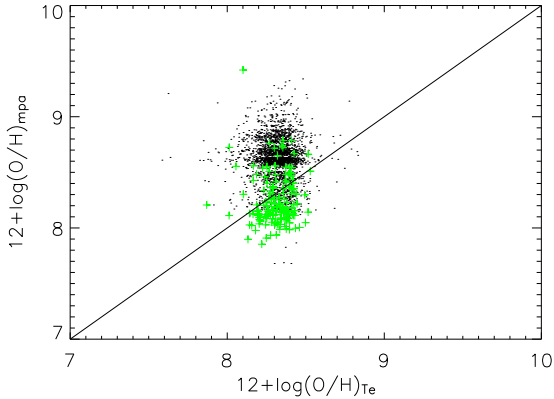}
\caption{Comparison of the $T_e$ metallcities and Bayesian metallcities.
Crosses are Sample I. Small dots are Sample II. }
\label{tempa}
\end{figure}

 In Fig.\ref{tempa}, we compare the $T_e$ and Bayesian 
metallicities in our SDSS sample. It shows that the Bayesian metallicities of 
nearly half  of the  {\bf Sample II}, where $T_e$ are derived from the empirical 
method in Pilyugin (2001), are $\sim$0.3dex higher 
than $T_e$ metallicity. The 
relationship between the $T_e$ and Bayesian metallicities 
 behaves very much like  the result in Yin \etal (2007) (their Fig.2),  which supports 
the empirical $T_e$ determination by Pilyugin (2001). This is 
consistent with $T_e$ derived by the classic temperature-sensitive line. 
Similar evidence can be found in Shi \etal (2006) (their Fig. 2) where 
$T_e$ metallicities from the empirical $T_e$ determination have nearly the same 
relation with other strong line metallicities as do those from $T_e$ derived from 
the classic temperature-sensitive line. 

 Generally speaking, all the methods should result in the same abundance  value for a given nebula, but this is not the case in  practice.
It is evident that this discrepancy is caused by the problems both
with models of HII regions and calibration. Why is the
calibration of some types, such as the $T_e$ method in this paper, 
regarded as the most reliable way to derive the oxygen abundance?

When comparing the numerical HII region models from
Charlot \& Longhetti (2001), which are the basis of the Bayesian 
abundances, with the numerical models of other authors 
(Stasi{\'n}ska \& Leitherer 1996; McGaugh 1991;  
CLOUDY by Ferland \etal 1998;  or Kewley \& Dopita 2002), 
one finds that there is  significant disagreement between 
them because these models use different atomic data or different
assumptions, and the stellar evolutionary  synthesis code and  
photoionization code used in these models continuously 
improved.  As a result,
 the present-day models cannot provide any uniform oxygen abundances.

As for calibrations, the calibrations based on the measurements 
of real \hii\ regions is believed to be the most favorable way to 
derive the oxygen abundances, because it is calibrated by single \hii\ region and
 will not be biased by other hot \hii\ regions. That is why we
 use empirical $T_e$ metallicity rather than Bayesian metallicity 
in high metallicity regions to calibrate $Ne3O2$ index. 
There is a problem that the amount of \hii\ 
regions with accurate measurements (a number of calibrating points)
are not numerous enough. Therefore one has to use extrapolation, which 
can be a cause of uncertainty in the abundance determination.

\begin{figure}
\centering
\includegraphics[angle=0,width=0.45\textwidth]{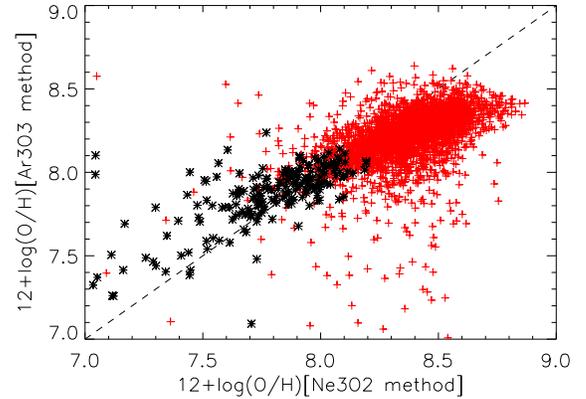}
\caption{
Comparison of the $Ne3O2$ and $Ar3O3$ methods. $Ar3O3$ metallicity  is 
calculated using the calibration of Stasi{\'n}ska (2006). $Ne3O2$ metallicity 
is  calculated using Eq. (2). Sample I is indicated by stars, and Sample II  is indicated by crosses.
}
\label{metal-indicator}
\end{figure}

Because $Ar$ and $Ne$ are produced by the same star, as expected by
 stellar nucleosynthesis (Izotov \& Thuan 1999), it is instructive to compare
 the $Ne3O2$--  and $Ar3O3$--methods (Stasi{\'n}ska 2006). We show this  
comparison for SDSS DR4 sample in Fig. \ref{metal-indicator}. 

Although  having a large scatter and a clear tendency of  the observed 
relation to slightly  deflect from  1:1, there is agreement
between the oxygen abundance from the $Ne3O2$-- method and the $Ar3O3$--method 
for most galaxies.  The dispersion between the $T_e$-based and 
 $Ar3O3$ methods  is especially large in the middle metallicity region
 ($7.95<\zoh<8.2$). It may be caused by the fact that, 
when calibrating $Ar3O3$ index, the oxygen abundance is calculated using
the empirical $P$--method, not the $T_e$-based method (Stasi{\'n}ska 2006), and  
the empirical $P$--method is problematic in the middle metallicity region,
 especially $7.95<\zoh<8.2$ (Pilyugin 2001).  The different oxygen abundance 
determination methods and different calibration methods
may be the cause of  the slope of the relation  slightly 
 deflecting from  1:1.   

The accuracy of the $Ar3O3$, $Ne3O2$ indicators are comparable, as can
 be judged from Fig. \ref{metal-indicator}. The advantage of the $Ne302$ method over the $Ar303$ one is that it does not demand a reliable reddening correction. The most prominent
 advantage of the $Ne3O2$ method over other metallicity indicators, such as
 the $R_{23}$-, $N2$-, $P$-, $O3N2$- methods, lies in its ability to determine 
the oxygen abundance for high redshift galaxies. The $Ne3O2$ line ratio can be 
measurable at the extreme redshift limit of ground-based optical surveys. 
The $Ne3O2$, $R_{23}$, $N2$, $P$, $O3N2$ line ratios can be detected within 
several NIR  atmospheric windows for the specific redshift z$\sim$1.5
 (Maier \etal 2006); but when it comes to the z$\sim$5-6 galaxy, only the $Ne3O2$ 
line ratio can be observed  with near-infrared instruments. When it comes to the
z$\sim$12 galaxy, the $Ne3O2$ line ratio can be detected with the James Webb Space
 Telescope (JWST), which uses a Near Infrared Camera and a Near Infrared 
Spectrometer (NIRSpec). 
\section{Conclusions}\label{conclusion}

 From the literature and SDSS DR4, we  collected  a large sample of
spectroscopic measurements of $\hii$ regions and $\hii$ galaxies covering 
a wide range in metallicity ($7.0<\zoh<9.0$). 
  The $T_e$ values for  210 galaxies (Sample I) in SDSS DR4 are  calculated with direct methods
 and   2960 galaxies (Sample II)   with an empirical method. We confirm the existence of  the correlation between $Ne3O2$ index and oxygen abundance
    and have obtained a calibration 
that can be used as a method for determining oxygen abundances. 
The dispersion between oxygen abundance and the  $Ne3O2$ 
index may come partly  from the moderate depletion of oxygen onto grains.

Though the $Ne3O2$ method is an empirical abundance determination method, we 
believe that also using the $Ne3O2$ index as a metallicity calibrator presents 
several advantages: 
\begin{enumerate}
\item
The $Ne3O2$ versus metallicity relation is monotonic.
\item
The relationship between $Ne3O2$ and metallicity is not affected by the 
chemical evolution effects because the Ne and O are produced by the same stars. 
\item
The $Ne3O2$ line ratio relies on the ratios of emission lines that are close 
in wavelength  so do not depend on reddening corrections. 
\item
This $Ne3O2$ line ratio can be measured for galaxies up to $z \sim 1.6$ using an  optical 
telescope, up to $z \sim 5.2$ using near-infrared instruments on the ground-based 
facilities, and up to $z \sim 12$ using JWST/NIRSpec; therefore, this flux 
ratio is a promising tool for metallicity studies at high redshift. 
\end{enumerate}

\begin{acknowledgements}
This work was supported by the Chinese National Science Foundation 
(No. 10521001, No. 10403006 and No. 10433010) and the National Basic Research Program of China (973 Program) No.2007CB815404. We thank G. Stasi{\'n}ska  and
 L.S. Pilyugin for their  
helpful comments and suggestions. We are grateful to the AGN 
group at the Center for Astrophysics, University of the 
Science of Technology of China for processing the SDSS spectra for continuum
decomposition and line fitting using the spectral analysis algorithm developed by
the group. Funding for the Sloan Digital Sky Survey (SDSS) has been provided
by the Alfred P. Sloan Foundation, the Participating Institutions, the National
Aeronautics and Space Administration, the National Science Foundation, the U.S.
Department of Energy, the Japanese Monbukagakusho, and the Max Planck Society.
The SDSS Web site is http://www.sdss.org/.  The SDSS is managed by the Astrophysical
Research Consortium (ARC) for the Participating Institutions. The Participating
Institutions are The University of Chicago, Fermilab, the Institute for Advanced
Study, the Japan Participation Group, The Johns Hopkins University, the Korean
Scientist Group, Los Alamos National Laboratory, the Max-Planck-Institute for
Astronomy (MPIA), the Max-Planck-Institute for Astrophysics (MPA), New Mexico
State University, University of Pittsburgh, University of Portsmouth, Princeton
University, the United States Naval Observatory, and the University of Washington.

\end{acknowledgements}

\end{document}